\documentclass[twocolumn,tighten,times]{aastex62}
\newcommand{\dPhi}{\ensuremath{\Delta \phi(r_0,r_{\rm tp})}}

\newcommand\teff{\ensuremath{T_{\rm eff}}}
\newcommand\logg{\ensuremath{\log\,g}}
\newcommand{\brv}{Brunt-V\"ais\"al\"a}
\newcommand\pcross{\ensuremath{P_{\rm cross}}}
\newcommand{\bvfreq}{Brunt-V\"ais\"al\"a frequency}
\newcommand\rtp{\ensuremath{r_{\rm tp}}}

\submitjournal{ApJ}
\received{2019 August 29}
\revised{2020 January 2}
\accepted{2020 January 8}

\shorttitle{Limits on Mode Coherence Due to a Non-static Convection
  Zone} 
\shortauthors{Montgomery, Hermes, Winget, Dunlap, and Bell} 

\begin{document}

\title{Limits on Mode Coherence in Pulsating DA White Dwarfs Due to a Non-static Convection Zone}

\author[0000-0002-6748-1748]{M. H. Montgomery}
\affiliation{University of Texas and McDonald Observatory, Austin,
  TX, USA}

\author{J. J. Hermes}
\affiliation{Department of Astronomy, Boston University, 725 Commonwealth Ave., Boston, MA 02215, US}

\author{D. E. Winget}
\affiliation{University of Texas and McDonald Observatory, Austin,
  TX, USA}

\author{B. H. Dunlap}
\affiliation{University of Texas and McDonald Observatory, Austin, TX, USA}

\author{K. J. Bell}
\affiliation{Department of Astronomy, University of Washington,
  Seattle, WA 98195-1580, USA}
\affiliation{NSF Astronomy and Astrophysics Postdoctoral Fellow}

\correspondingauthor{M. H. Montgomery}
\email{mikemon@astro.as.utexas.edu}

\begin{abstract}

  The standard theory of pulsations deals with the frequencies and
  growth rates of infinitesimal perturbations in a stellar
  model. Modes which are calculated to be linearly driven should
  increase their amplitudes exponentially with time; the fact that
  nearly constant amplitudes are usually observed is evidence that
  nonlinear mechanisms inhibit the growth of finite amplitude
  pulsations. Models predict that the mass of convection zones in
  pulsating hydrogen-atmosphere (DAV) white dwarfs is very sensitive
  to temperature (i.e., $M_{\rm CZ} \propto T_{\rm eff}^{-90}$),
  leading to the possibility that even low-amplitude pulsators may
  experience significant nonlinear effects.  In particular, the outer
  turning point of finite-amplitude g-mode pulsations can vary with
  the local surface temperature, producing a reflected wave that is
  out of phase with what is required for a standing
  wave. This can lead to a lack of coherence of the mode and a
  reduction in its global amplitude. In this paper we show that: (1)
  whether a mode is calculated to propagate to the base of the
  convection zone is an accurate predictor of its width in the Fourier
  spectrum, (2) the phase shifts produced by reflection from the outer
  turning point are large enough to produce significant damping, and
  (3) amplitudes and periods are predicted to increase from the blue
  edge to the middle of the instability strip, and subsequently
  decrease as the red edge is approached. This amplitude decrease is
  in agreement with the observational data while the period decrease
  has not yet been systematically studied.


\end{abstract}


\section{Astrophysical Context}

In the linear, adiabatic theory of stellar pulsations, modes are
considered to be perfectly sinusoidal in time. This results in a
theoretical eigenmode spectrum with arbitrarily thin, delta function
peaks as a function of frequency. In the linear, \emph{non}-adiabatic
theory, modes are allowed to gain or lose energy with their
environment, leading to non-zero growth/damping rates, and these rates
may be related to the finite widths of peaks in their power
spectra. In particular, the widths of modes in solar-like pulsators
can be linked to their linear damping rates
\citep[e.g.,][]{Houdek15,Kumar89}.

In the white dwarf (WD) regime, we have direct observational evidence
that some modes in pulsating white dwarfs can show a high degree of
phase coherence, in a few cases spanning decades. For instance, more
than 40 years of observations have established that the phase of the
215~s mode in the pulsating hydrogen-atmosphere white dwarf (DAV)
G117-B15A is coherent over this time scale. Thus, we can show that in
this DAV the pulsation period, $P$, is changing (taking into account
the proper motion) at the extremely slow rate of
$\dot{P} \rm = (3.57 \pm 0.82) \times 10^{-15} s\,s^{-1}$
\citep{Kepler05a}. The extreme sensitivity of such $\dot{P}$
measurements in this and other white dwarfs has allowed them to be
used as testbeds for physical processes that could affect
their cooling, such as the interaction of hypothetical dark matter
particles
\citep[e.g.,][]{Isern08,Kim08b,Corsico12a,Corsico16}.

Another use of the observed stability of these modes is searching for
planetary signals in the delayed and advanced light arrival times due
to reflex orbital motion of the white dwarf
\citep{Mullally03,Mullally08,Hermes10,Winget15}, although no planets
have been positively identified orbiting WDs with this technique to
date. We note that while the coherent modes that are useful for these
studies are mostly found in DAVs near the hot edge of the instability
strip, striking seasonal changes in the Fourier spectra of cooler DAVs
are commonly observed \citep[e.g.,][]{Kleinman98}. The transition from
pulsators with stable Fourier spectra near the blue edge to those
showing frequent amplitude changes near the red edge occurs somewhere
in the middle of the observed DAV instability strip, in the vicinity
of $\teff \sim\,$11,500~K.

While a handful of coherent modes have been studied in DAVs, no
systematic study of mode coherence in a large sample of these stars
has been made from the ground.  This situation has improved greatly
with the launch of the \emph{Kepler} spacecraft. During its original
mission and the follow-on \emph{K2} mission, it has obtained
nearly-continuous time series data, often exceeding 75 d, of a large
number of pulsating white dwarf stars. Recently, \citet{Hermes17a}
published comprehensive data on the first 27 DAVs studied by
\emph{Kepler}. One of their central results was that longer-period
modes ($P \gtrsim 800$~s) were observed to have larger Fourier widths
than shorter-period modes ($P \lesssim 800$~s), essentially dividing
the modes into two populations; this result can hold even for
different modes in the \emph{same} star. We present an explanation
for this phenomenon based on whether the mode propagates to the base
of the non-stationary convection zone. We also show that this leads
to damping of the modes and that the effect is larger near the red
edge of the DAV instability strip.


\begin{figure}[t]
  \centering{\includegraphics[width=\columnwidth]{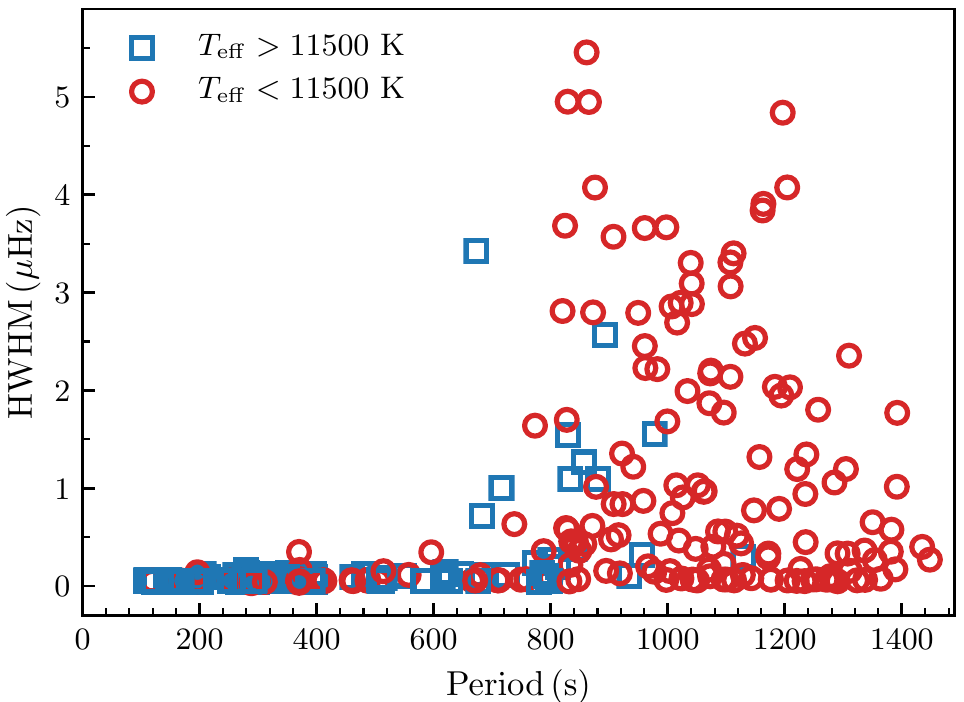}} 
  \caption{ Observed Fourier width versus period for modes in DAVs
    observed with \emph{Kepler} and \emph{K2}. The blue squares and
    red circles denote modes in stars with $\teff > 11$,500~K and
    $\teff < 11$,500~K, respectively.  }
\label{hermes}
\end{figure}

\section{The Data}

The extended length of observations
($\gtrsim$$75$~d for most stars) in the \emph{Kepler} and \emph{K2}
data sets results in a $1/T$ resolution in the power spectra of $<$$\,
0.14 \,\, \mu \rm Hz$; this sets the observable lower limit for the
width of a peak in the Fourier transform. For the first time this
enables the measurement of the widths of a large number of modes in
many stars that are above this threshold \citep{Bell15a}. For their
sample of 27 DAVs, \citet{Hermes17a} obtained follow-up spectroscopy
with the WHT and SOAR telescopes; these spectra were fit using the
techniques and models of \citet{Tremblay11a} to obtain values of
\teff\ and \logg\ for each star.

\citet{Hermes17a} find that the Fourier width of modes, (HWHM, the
half width at half maximum of a Lorentzian fit), is a strong function
of the mode period, $P$. To summarize, they find that 1) modes with
$\rm HWHM > 0.3\,\mu$Hz have $P \gtrsim 800$~s and 2) modes with
$P \lesssim 800$~s have $\rm HWHM < 0.3\,\mu$Hz.  This is illustrated
in Figure~\ref{hermes}, in which we have plotted mode width versus
period for the linearly independent periods found in the sample of
DAVs from \citet{Hermes17a}\footnote{We have revisited the mode
  identification from \citet{Hermes17a} and believe that two outliers
  in Figure 1 that correspond to $f_6$ and $f_8$ of EPIC\,210397465
  are not independent modes but are most likely nonlinear combination
  frequencies (where $f_6 = f_{3b}+f_5$ and $f_8 = f_{3a}+f_5$). We
  therefore do not include these nonlinear combination frequencies in
  our analysis here.}.
Modes from stars with $\teff > 11$,500~K are shown as blue squares,
while those from stars with $\teff < 11$,500~K are shown as red
circles. The fact that most of the modes with $\rm HWHM > 0.3\,\mu$Hz
are from the cooler population is not an independent piece of
information since longer-period modes are typically found only in the
cooler DAVs \citep{Mukadam06b}.



\begin{figure}[t]
  \centerline{\includegraphics[width=1.0\columnwidth]{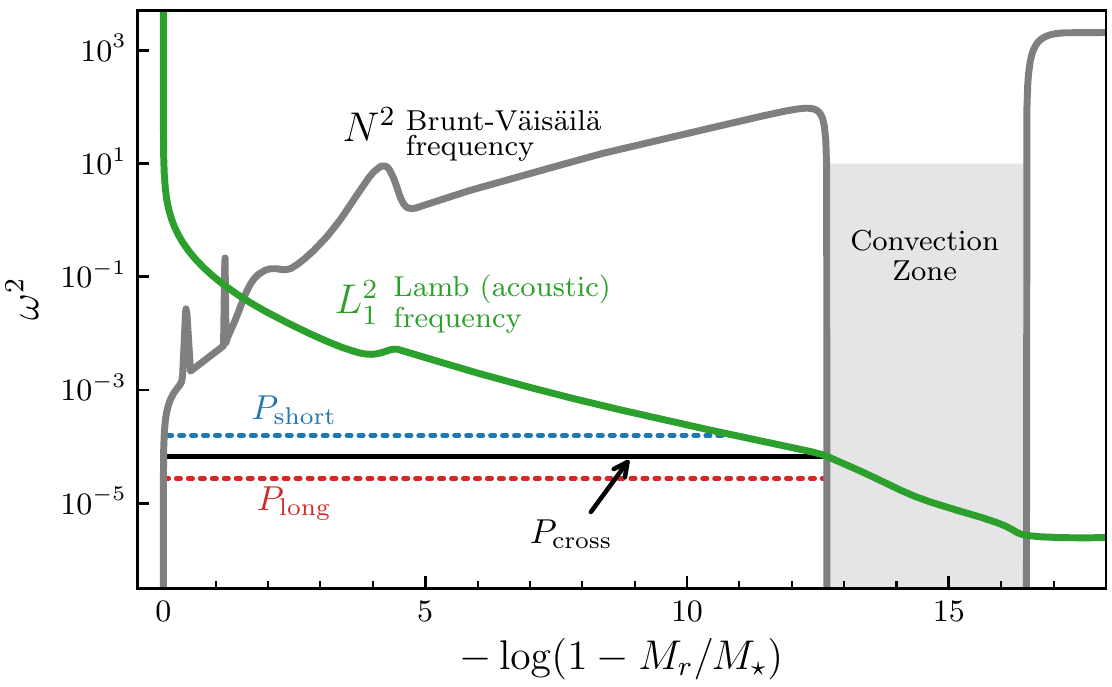}}
  \caption{Schematic propagation diagram of a DAV white dwarf.  A
    g-mode with a ``short'' period would have an outer turning point
    beneath the convection zone (blue dotted line), while one with a
    ``long'' period would have an outer turning point at the
    convection zone boundary (red dotted line). The crossover point
    between these two regimes is given by a mode with
    $P=P_{\rm cross}$ (black horizontal line).  }
  \label{prop1}
\end{figure}

\begin{figure}[t!]
  \centerline{\includegraphics[width=1.0\columnwidth]{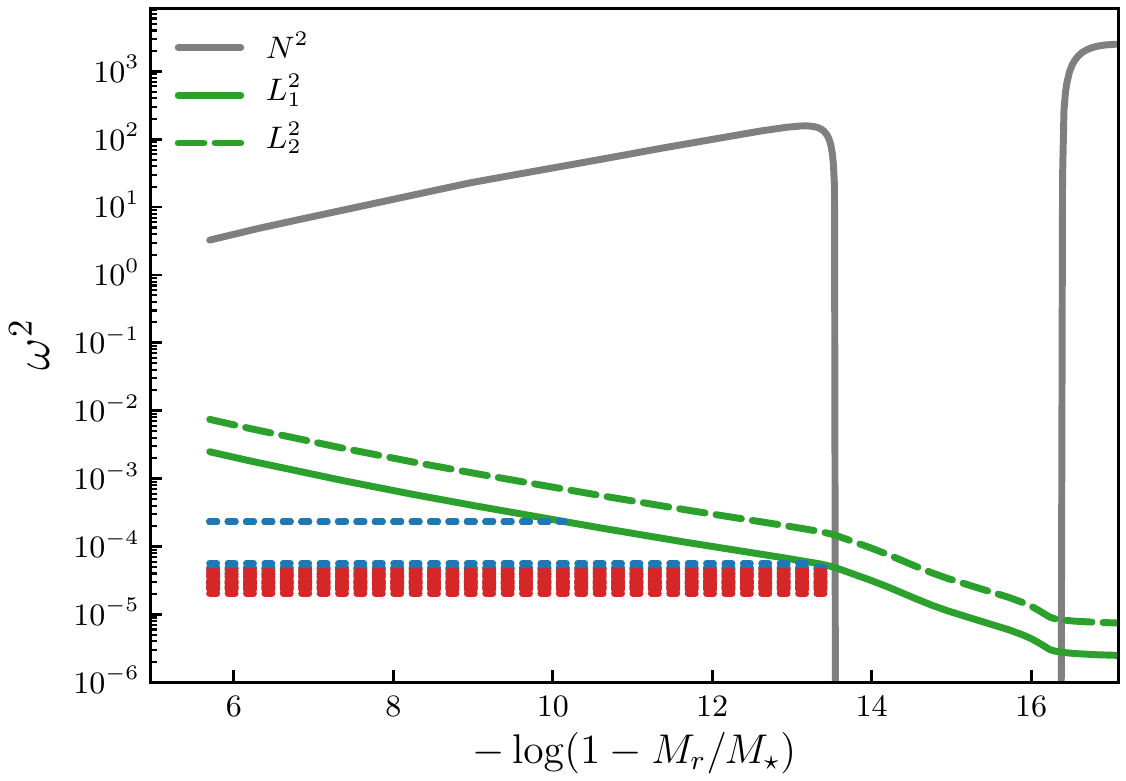}}
  \caption{The outer propagation region for a model with the same
    parameters as EPIC 201806008. The dashed horizontal lines show the
    region of propagation that the observed modes would have in this
    model: the blue dashed lines denote modes observed to have
    $\rm HWHM < 0.3\, \mu$Hz and the red dashed lines modes with
    $\rm HWHM > 0.3\, \mu$Hz. The solid and dashed lines show the run of the Lamb (acoustic) frequency for $\ell=1$ and $\ell=2$ modes, respectively, to illustrate that higher-$\ell$ modes have lower values of $P_{\rm cross}$.}

  \label{epic1}
\end{figure}

\section{The Propagation Region}

The region of propagation of $g$ modes (``gravity'' or ``buoyancy''
modes) in a star is defined by the region in which
$\omega^2 < N^2, L^2_{\ell}$,
where $\omega = 2 \pi/P$ is the angular frequency of the mode,
$N$ is the \brv\ (buoyancy) frequency,
$L_{\ell}$ is the Lamb (acoustic) frequency, and
$\ell$ is the spherical degree of the mode. In Figure~\ref{prop1}, we
show a propagation diagram for a model, computed with MESA
\citep{Paxton11,Paxton13,Paxton15,Paxton18,Paxton19}.  The black
horizontal line denotes the region of propagation of a hypothetical
mode whose period, \pcross, is the minimum required for it to
propagate to the base of the convection zone.

In Figure~\ref{epic1}, we show the outer propagation region for a
model with the same parameters as EPIC 201806008 ($\log
\,g=8.02$, $\teff =
10910\,$K).  The dashed horizontal lines show the region of
propagation that the observed modes would have in this model, where
the blue dashed lines denote modes observed to have $\rm HWHM < 0.3\,
\mu$Hz and the red dashed lines modes with $\rm HWHM > 0.3\,
\mu$Hz.  We note that, by using the convection model $\rm ML2/\alpha =
0.42$, where
$\alpha$ is the mixing length to pressure scale height ratio
\citep[see][]{Bohm71}, we can divide the modes into two groups: 1)
narrow (blue) modes whose outer turning point is well beneath the
convection zone, and 2) wide (red) modes that propagate all the way to
the base of the convection zone. In other words, all the modes with
$\rm HWHM >
0.3\,\mu$Hz can be explained as propagating to the base of the
convection zone, whereas all the modes with $\rm HWHM <
0.3\,\mu$Hz have an outer turning point safely inside this point.
{\bf Our hypothesis is that the long-period modes, through their
  interaction with the convection zone, will have systematically
  larger Fourier mode widths than the short-period modes.} In the
following sections we examine this statement more quantitatively.

\begin{figure}[t!]
  \centerline{\includegraphics[width=1.0\columnwidth]{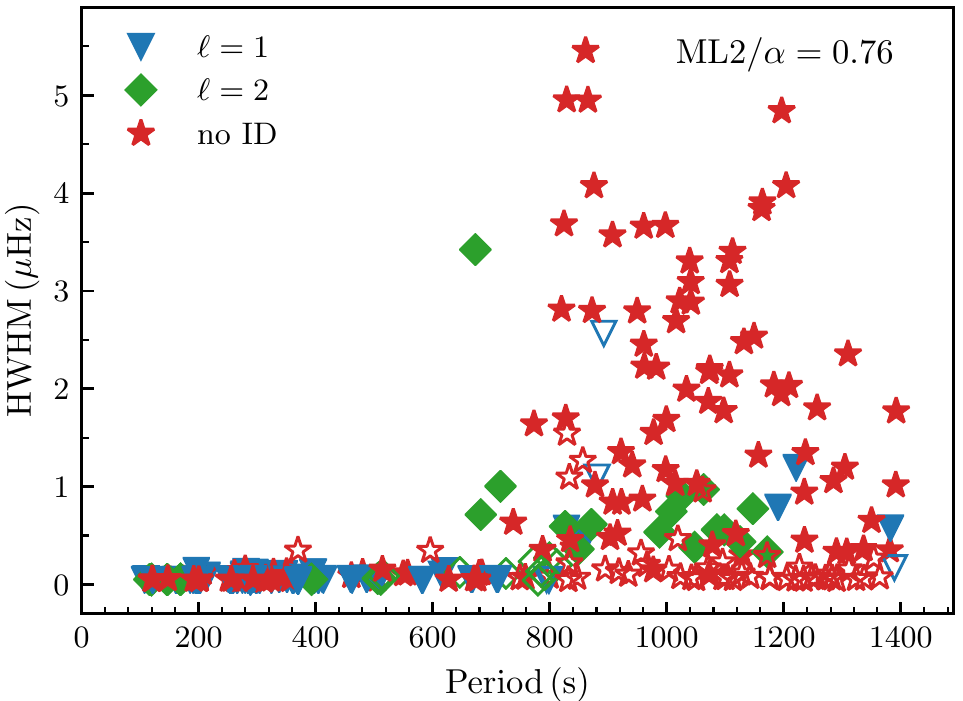}}
  \caption{Similar to Figure~\ref{hermes}, but with the following
    differences: blue triangles, green diamonds, and red stars denote
    modes that are identified as $\ell=1$, $\ell=2$, or unidentified
    modes, respectively. Using \pcross\ for each star as
    computed from its \teff\ and \logg\ values, modes with
    $P \ge \pcross$ and $\rm HWHM > 0.3 \,\mu$Hz are shown as filled
    symbols, as are modes with $P < \pcross$ and
    $\rm HWHM < 0.3\, \mu$Hz.  The opposite cases ($P < \pcross$ and
    $\rm HWHM > 0.3 \,\mu$Hz, or $P \ge \pcross$ and
    $\rm HWHM < 0.3\, \mu$Hz) are shown as unfilled symbols.  }
  \label{predict}
\end{figure}

\section{A First Observational Test}

Since the 27 DAVs in \citet{Hermes17a} have different atmospheric
parameters, each star will have a different value of \pcross\ (note:
\pcross\ is also a function of the $\ell$ value of each mode). Thus,
for each mode in each star we calculate \pcross\ and from this we
predict whether each observed mode in the star has a ``wide, mottled''
peak in the FT or a ``narrow'' peak. We assume the ML2 convection
model of \citet{Bohm71} with $\rm ML2/\alpha=0.76$. This value of
$\alpha$ is taken from the 3D simulations of \citet{Tremblay15a} for a
model with $\teff =11,500$~K and $\logg = 8.0$.

We plot the results of this in Figure~\ref{predict}, where we have
used different symbols for the different $\ell$ identifications: blue
triangles for $\ell=1$ modes, green diamonds for $\ell=2$ modes, and
red stars for unidentified modes.  Unidentified modes are assumed to
have $\ell=1$ for this analysis. Modes with wide peaks (i.e.,
$\rm HWHM > 0.3\,\mu$Hz) are predicted to interact strongly with the
convection zone. If this prediction is correct (i.e., if
$P \ge \pcross$ for the mode) then they are plotted with a filled
symbol. If the prediction is incorrect, they are plotted with an
unfilled symbol.  Similarly, modes with narrow peaks (i.e.,
$\rm HWHM < 0.3\,\mu$Hz) are predicted to interact weakly with the
convection zone. If this prediction is correct (i.e., if
$P < \pcross$) then they are also plotted with a filled symbol;
otherwise, an unfilled symbol is used. Thus, the filled symbols
represent the modes whose widths are correctly predicted by our
hypothesis. We find that this procedure correctly classifies the
observed mode widths with an accuracy of $\sim\,$81\%. While there is
evidence that the value of $\alpha$ is a function of \teff\
\citep{Tremblay15a,Provencal12}, the percentage of correctly predicted
modes is roughly constant for $\alpha$ in the range 0.6--0.9.

\begin{figure}[t]
  \centering{
  \includegraphics[width=0.985\columnwidth]{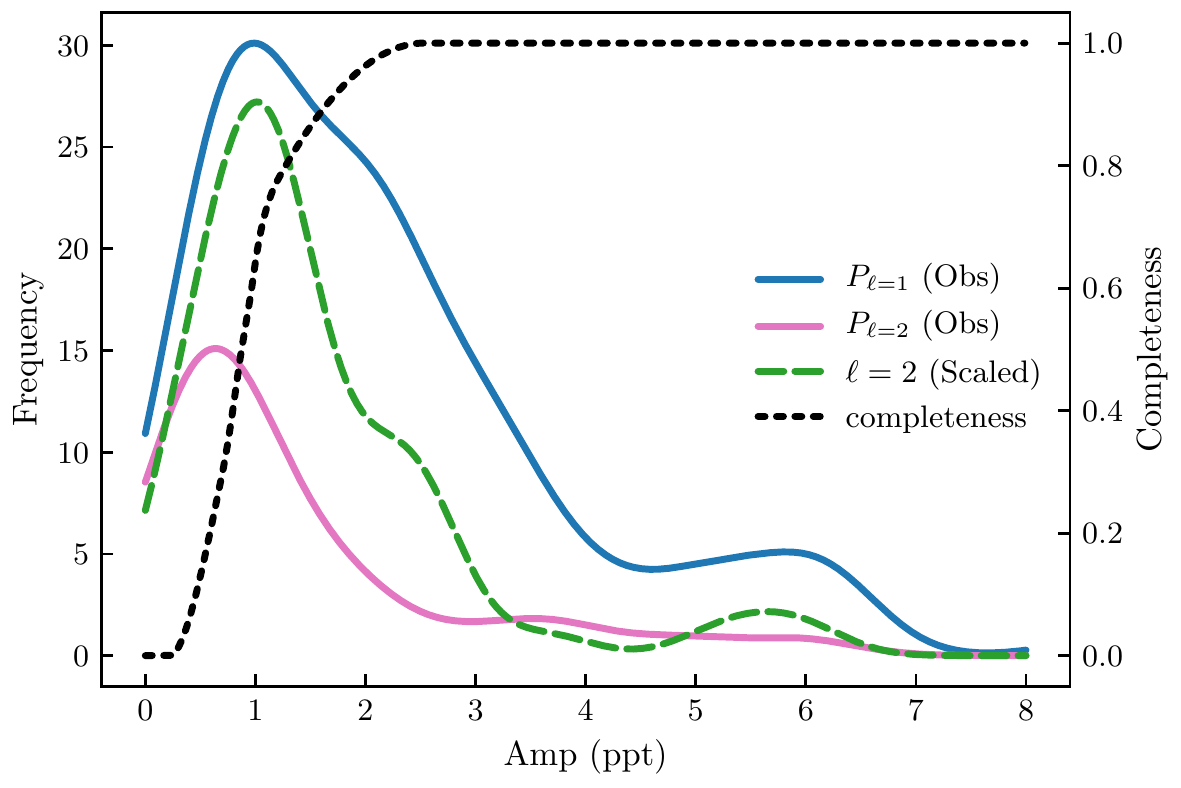}
  }
  \vspace*{-1em}
  \caption{The observed distribution of modes identified as $\ell=1$
    and 2 modes ($D_{\ell=1}$, $D_{\ell=2}$; blue and pink curves,
    respectively) as a function of amplitude. The $\ell=2$ (Scaled)
    curve (green dashed) shows the distribution that is obtained by
    scaling the $\ell=1$ distribution by the expected geometric
    amplitude ratio of $\ell=2$ and $\ell=1$ modes. We note that this
    results in a greater number of $\ell=2$ modes is observed.  We
    note that the different detection thresholds of each star have
    been folded into a ``completeness'' curve, whose values are shown
    on the right vertical axis.  }
  \label{amp_dist}
\end{figure}

\section{A Second Observational Test}

Modes of different spherical degree ($\ell$) are also expected to
interact with the convection zone differently as a function of period.
Since the Lamb frequency of $\ell=2$ modes is larger than it is for
$\ell=1$, for a fixed frequency $\ell=2$ modes have an outer turning
point that is closer to the stellar surface, and are therefore more
likely to strongly interact with the convection zone. If this can
lead, in some cases, to not just a broadened peak in the FT but to a
partial or complete suppression of the mode, then we would expect the
number of $\ell=2$ modes to be suppressed by this mechanism.

\citet{Hermes17a}
classify all of the statistically significant peaks in the FTs as
either $\ell=1$, $\ell=2$, or undetermined.  They were able to
identify the $\ell$ value of 87 out of 201 independent
modes in these DAVs (more than 40\%), based on common frequency
patterns in the Fourier transforms (e.g., the rotational splitting of
multiplets and the period spacing of different radial overtones).

In Figure~\ref{amp_dist}, we use ``kernel density estimation'' (kde)
to estimate the density of modes as a function of the heights of the Lorentzians that were fit to them in the power spectrum, which we call their amplitude
\citep{Pedregosa11}. Only modes containing peaks with less than a 0.1\% false alarm
probability (FAP) and whose frequencies are not linear combinations
of other mode frequencies are considered. As expected, the $\ell=1$
modes (blue curve) outnumber $\ell=2$ modes (pink curve).  We next
wish to test the hypothesis that there are as many $\ell=2$ as
$\ell=1$ modes, but that the difference in observed numbers is solely
due to geometric cancellation reducing their observed amplitudes below
detection limits.

To test this, we divide the observed $\ell=1$ amplitudes by a factor
of 2.4 to simulate the larger geometric cancellation of $\ell=2$ modes
observed in the \emph{Kepler} passband.  The result is shown as the
green dashed curve in Figure~\ref{amp_dist}.  We see that the number
of $\ell=2$ modes obtained in this way exceeds those observed by a
factor of more than 2.  This could indicate that a mechanism in
addition to pure geometric cancellation is limiting their number. Of
course, without a way of identifying the unidentified modes that
comprise $\sim\,60$\% of this sample, it is not possible to
conclusively demonstrate this.

Finally, we note that after scaling the observed $\ell=1$ amplitudes
by the appropriate factors for $\ell = 3$ and 4 ($\sim 26$ and 19,
respectively), no modes are found to be above the $\rm FAP > 0.1$\%
detection threshold. Thus, these data have nothing to say on the
possible presence or amplitude distribution of these higher-$\ell$
modes.

\section{Interaction with the Convection Zone}
\label{interaction}

\begin{figure}[t!]
  \centerline{\includegraphics[width=1.0\columnwidth]{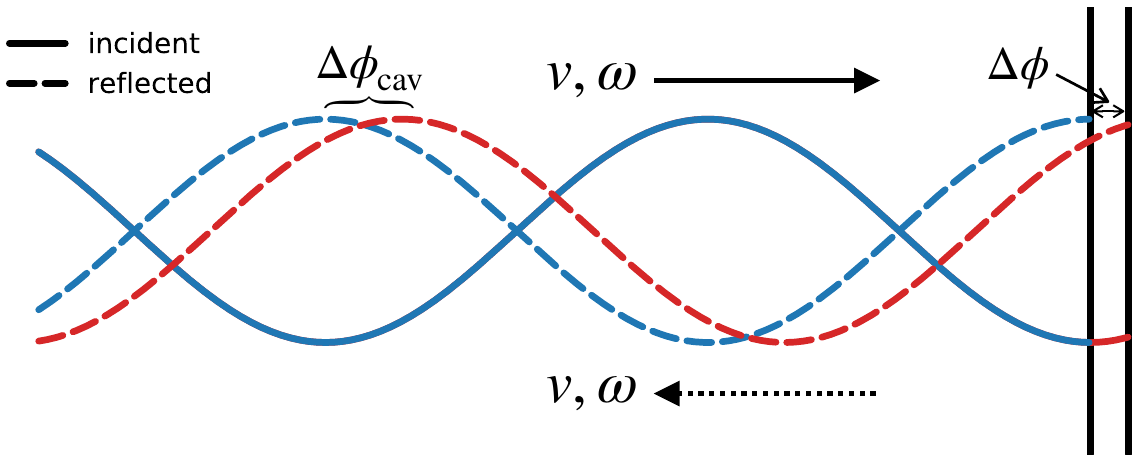}}
  \caption{\emph{Solid lines}: the incoming wave incident on the base
    of the convection zone (vertical black lines); \emph{dashed
      lines}: the waves reflected from the base of the convection
    zone. The blue curves represent the unperturbed case (reflection
    from leftmost vertical line) while the red curves show the effect
    of moving the base of the convection zone to the right (rightmost
    vertical line). The phase difference of the maxima of the
    reflected waves in these two cases is labeled as
    $\Delta \phi_{\rm cav}$, which is twice the phase shift
    $\Delta \phi$ labeled on the rightmost boundary.  }
  \label{ref1}
\end{figure}

\begin{figure}[t!]
  \centerline{\includegraphics[width=1.0\columnwidth]{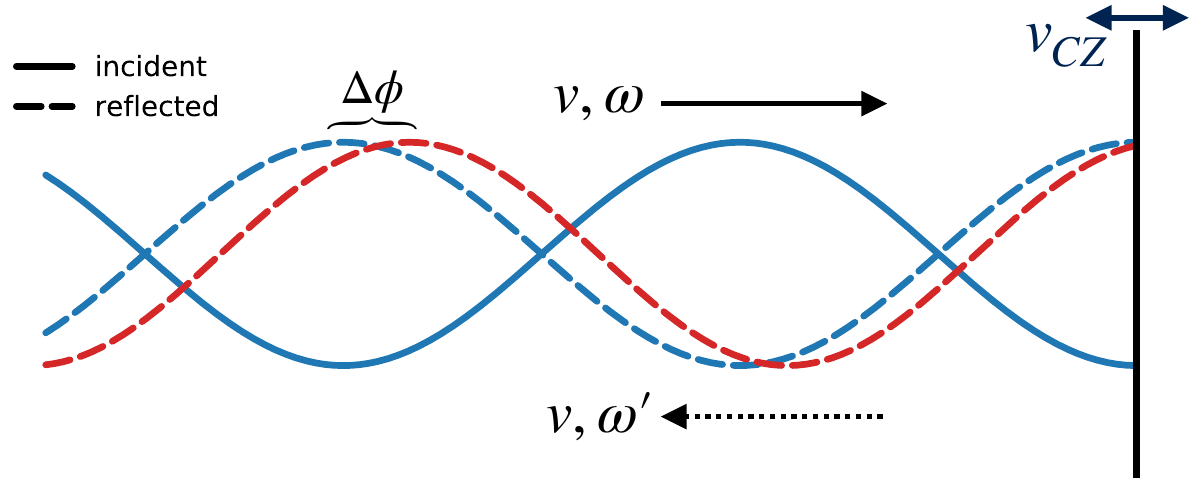}}
  \caption{The change in the frequency of the reflected wave
    ($\omega \rightarrow \omega'$) due to the velocity of the base of
    the convection zone ($v_{CZ}$).  The blue curves represent
      the unperturbed case (reflection from a stationary boundary)
      while the red curve shows the effect of a convection zone base
      moving to the left. This frequency change leads to a change in
    the radial wavenumber of the wave, leading to a slow accumulation
    of phase difference as the wave propagates.  }
  \label{ref2}
\end{figure}

There are two main ways that a mode's interaction with the outer
convection zone could lead to a loss of its coherence. First, if a
mode has a low enough frequency, its outer turning point will be the
base of the convection zone.  Since the base of the convection zone
rises and falls with the surface temperature perturbations of the
pulsations, the mode will sometimes have to travel a greater (or
lesser) distance before being reflected, and in so doing it will
acquire an extra phase $\Delta \phi$ (see Figure~\ref{ref1}).  A
second effect that may play an even larger role is the Doppler shift
of the wave caused by reflection from the base of the moving
convection zone (see Figure~\ref{ref2}).  Essentially, the reflection
causes a frequency shift of the reflected wave, which again leads to a
phase mismatch when the wave next returns to the outer turning point.

In \citet{Montgomery15b}, we showed that the extra phase acquired by a
mode with quantum numbers $\{n,\ell\}$ due to a change in the size of
its propagation cavity could be approximately calculated from the
difference in period of this mode in two models ($\Delta P$) that
differ \emph{only} in the depth of their convection zones; this extra
phase is given by
\begin{equation}
  \Delta \phi_{\rm cav} = 2 \pi n \left(\frac{\Delta P}{P}\right).
  \label{cav}
\end{equation}

We also calculate this phase difference by directly integrating the
asymptotic expression for the radial wavenumber. From
  \citet{Gough93} we have that the asymptotic radial wavenumber $K(r)$
  is given by
\begin{equation}
  K^2(r) = \frac{\omega^2-\omega_c^2}{c^2} -\frac{L^2}{r^2} \left( 1 -
    \frac{N^2}{\omega^2}\right),
  \label{Kgough}
\end{equation}
where $N$ is the \bvfreq, $L^2 \equiv \ell (\ell+1)$, $c$ is the sound
speed, $\omega$ is the angular frequency of the mode, and $\omega_c$
is the acoustic cutoff frequency. The phase difference is then
\begin{eqnarray}
  \Delta \phi_{\rm cav} & = & 2 \,\Delta \phi(r_0,\rtp) \nonumber \\
& \equiv & 2 \left[ \int_{r_0}^{\rtp} K(r') dr' - \int_{r_0}^{r_{\rm t p
      0}} K_0(r') dr' \right],
\label{dphi}
\end{eqnarray}
where $r_0$ is a fiducial lower radius that remains fixed in mass and
\rtp\ is the outer turning point of the mode; we note that \rtp\ is
the actual radius at which reflection occurs (i.e.,
$K(r_{\rtp}) = 0$).  The second term in brackets on the RHS of
equation~\ref{dphi} is the equilibrium value of the phase, and the
first term is its instantaneous value. The factor of 2 is to account
for the fact that the wave propagates from $r_0$ to \rtp, and then
back to $r_0$.  For long period modes the reflection point is very
near the base of the convection zone, but for shorter period modes
\rtp\ is deeper in the model.

As previously mentioned, the Doppler shift caused by reflection
from the base of the moving convection zone may lead to even larger
phase shifts (see Figure~\ref{ref2}).  Essentially, the frequency
shift due to reflection leads to a change in the wavelength of the
wave \citep{Montgomery19a}. Thus, in traveling down to the inner
turning point and back the mode no longer travels an exact integer
number of wavelengths, arriving back at the outer turning point with a
phase shift. Formally, the frequency perturbation is given by
\begin{equation}
  \omega' = \omega \, \left( 1 - v_{CZ}/v_{\rm gr} \right),
  \label{dopa}
\end{equation}
or, setting $\omega' = \omega + \Delta \omega$,
\begin{equation}
  \frac{\Delta \omega}{\omega} = - \frac{v_{CZ}}{v_{\rm gr}},
  \label{dop}
\end{equation}
where $\omega$ is the frequency of the incident wave, $v_{CZ}$ is the
velocity of the base of the convection zone, and $v_{\rm gr}$ is the
group velocity of the wave at the base of the convection zone, given
by $v_{\rm gr} \equiv \partial \omega/\partial K$.

Equation~\ref{dop} is non-trivial to evaluate since $v_{\rm gr}$
formally goes to zero near the base of the convection zone.
To remove this difficulty, we imagine the
reflection occurring from a surface of constant phase well beneath the
outer turning point, \rtp, and we calculate the velocity of this
surface, $v_{\rm surf}$. Using asymptotic formulae, we write this
phase between the radii $r$ and \rtp\ as
\begin{eqnarray}
  \phi & \equiv & \int_r^{\rtp} K(r') dr' \\
  & = & \int_r^{r_{0}} K(r') dr'  + \int_{r_0}^{\rtp} K(r') dr',
\label{phi}
\end{eqnarray}
where $r$ is the radius of the surface of constant phase, $K(r)$ is given
by equation~\ref{Kgough}, and $r_0$ is a fiducial radius between $r$
and \rtp\ whose value is held constant. 


Setting $\partial \phi/\partial t = 0$ in equation~\ref{phi} and
solving for $\partial r/\partial t$ yields
\begin{equation}
       \frac{\partial r}{\partial t} = \frac{1}{K(r)}
       \frac{\partial}{\partial t}
       \int_{r_0}^{\rtp} K(r') dr'.
\end{equation}
Since $\partial r/\partial t$ represents the velocity of the surface
of constant phase, we set $v_{\rm surf} =\partial r/\partial t$. Using
this in place of $v_{CZ}$ in equation~\ref{dop}, we find
\begin{eqnarray}
  \frac{\Delta \omega}{\omega} & = & - \frac{v_{\rm surf}}{v_{\rm gr}} \\
  & = & - \frac{1}{v_{\rm gr} K(r)} \frac{\partial}{\partial t}
       \int_{r_0}^{\rtp} K(r') dr'.
  \label{dopb}
\end{eqnarray}
Since we take the point $r$ to be safely beneath the outer turning
point, we can use the asymptotic expression for the group velocity of
g-modes, $v_{\rm gr} \approx - \omega/K(r)$. Substituting this in
equation~\ref{dopb} yields
\begin{equation}
  \Delta \omega = \frac{\partial}{\partial t} \int_{r_0}^{\rtp} K(r') dr'.
  \label{domdop}
\end{equation}
We recognize the integral on the RHS of equation~\ref{domdop} as the
mode phase between the point $r_0$ and the outer turning
point. Since the time derivative only acts on the time-dependent
variations of this quantity, we can rewrite it using
$\Delta \phi(r_0,\rtp)$ as defined in equation~\ref{dphi}. Thus,
\begin{eqnarray}
  \Delta \omega
  & = & \frac{\partial}{\partial t} \left[ \Delta \phi(r_0,\rtp)
        \right] \nonumber \\
  & \approx & \langle \omega \rangle \, \Delta \phi(r_0,\rtp),
  \label{domdop2}
\end{eqnarray}
where we have replaced the time derivative with the factor
$\langle \omega \rangle = 2\pi/\langle P \rangle$, where
$\langle P \rangle$ is a characteristic pulsation time scale for the
ensemble of excited modes in the star (e.g., shorter for hotter stars,
longer for cooler stars). Based on the results of \citet{Mukadam06b},
we take $\langle P \rangle = 300$~s for $\teff = 12,000$~K,
$\langle P \rangle = 800$~s for $\teff = 11,500$~K, and
$\langle P \rangle = 1000$~s for $\teff \le 11,000$~K.

We note that the large time-dependent changes occur near the base of
the convection zone, so $\Delta \phi(r_0,\rtp)$ and therefore
$\Delta \omega$ should be independent of the reference depth $r_0$.
Numerically, we find that $\Delta \phi(r_0,\rtp)$ changes by
$< 0.01$\% as $r_0$ increases in depth by one mode wavelength.  This
is necessary since the calculated value of the frequency shift
$\Delta \omega$ should be independent of the assumed distance from the
base of the convection zone at which the calculation is made.  For the
calculations in Section~\ref{res}, we arbitrarily choose this fiducial
radius to be the point below the outer turning point at which
$\phi = 5\pi/4$; this is slightly over half a mode wavelength beneath
the outer turning point.

The frequency difference given by equation~\ref{domdop2}, as the ray
propagates down to the inner turning point and back out to the outer
turning point, results in a total accumulated phase change of
\begin{eqnarray*}
\Delta \phi_{\rm dop} & = & 2 \pi n \left(\frac{\Delta
                            \omega}{\omega}\right) \\
          & = & 2 \pi n \,\frac{\langle \omega \rangle}{\omega} \,\, \dPhi
                            \label{dop2}
\end{eqnarray*}
These phase shifts lead to damping of the mode, as we show in the
following section. Finally, using equation~\ref{dphi} we see that 
the phase change due to the doppler shift of the mode frequency can be
related to that due to the changing size of the propagation cavity,
i.e., 
\begin{equation}
  \Delta \phi_{\rm dop} = n \pi \, \frac{\langle \omega \rangle}
    {\omega} \, \Delta \phi_{\rm cav}.
\end{equation}
Since $n \ge 1$ and $\omega \approx \langle \omega \rangle$, we expect
that the phase shift (and therefore the damping rate) due to the
doppler shift to dominate that due to the changing cavity size.

We note here that this approach assumes that the traveling waves that
comprise the mode propagate freely from the outer turning point down
to the inner turning point and back again without encountering any
regions of rapid spatial variation. Such regions could produce partial
reflection and transmission of the waves, resulting in an unequal
distribution of kinetic energy between the core and envelope regions.
This ``mode trapping'' would lead to differential sensitivity of
consecutive radial orders to phase shift effects. The fact that we
ignore these possible internal reflections means that mode
  trapping effects will be suppressed in our damping calculations.

\section{Theoretical Damping Rates}
\label{drates}

For the mode to be completely coherent, it needs to accumulate exactly
$2 \pi n$ radians of phase each time it propagates back and forth in
the star. As shown in the previous section, a changing convection zone
can upset this condition, leading to destructive interference and a
change in amplitude given by
\begin{equation}
  \frac{dA}{dt} = - \frac{A}{n P} \left(1-\cos \Delta \phi\right)
\end{equation}
\citep[see][]{Montgomery15b}. Assuming that
$A \propto e^{-\gamma\, t}$, and averaging $\gamma$ over values of the
phase shift from $-\Delta \phi$ to $+\Delta \phi$ gives
\begin{equation}
  \gamma =\frac{1}{n P} \left( 1 -
    \frac{\sin \Delta \phi}{\Delta \phi} \right).
  \label{gamma}
\end{equation}
This is the equation we use to calculate the damping rates of modes
given the amplitudes ($\Delta \phi$) of their phase shifts.

\begin{figure}[t!]
  \centerline{
    \includegraphics[width=\columnwidth]{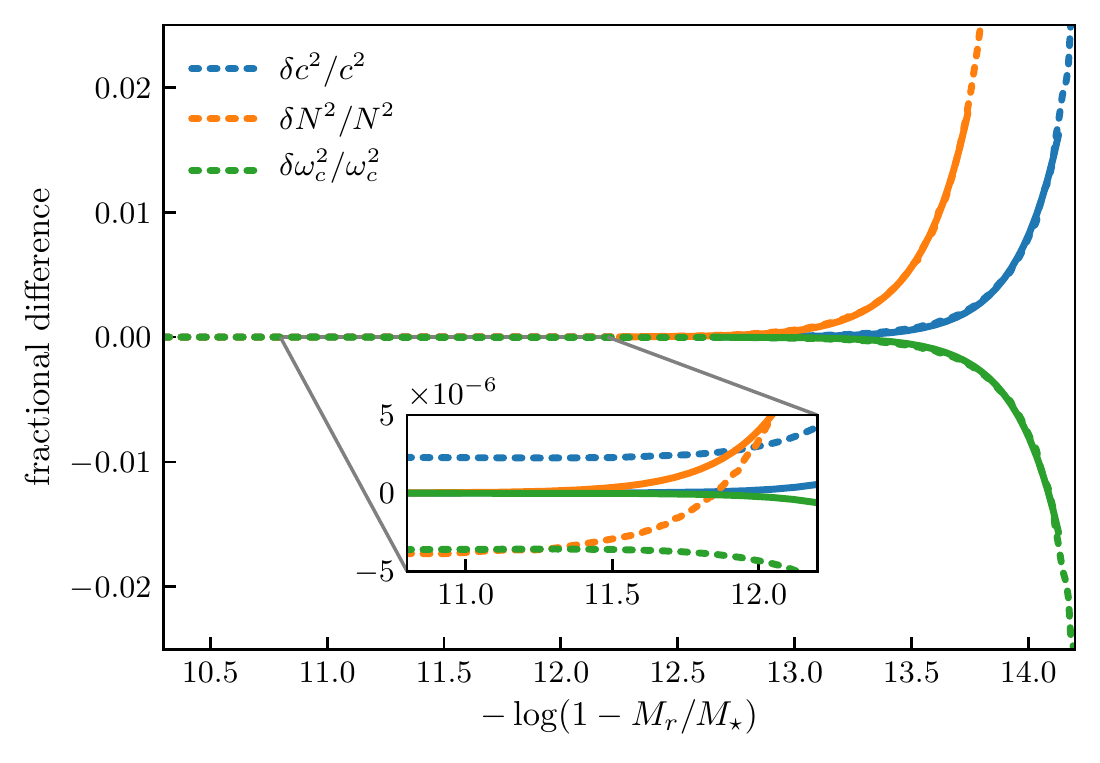}
  }
  \caption{Difference of the pulsation quantities $N^2$, $c^2$,
    and $\omega_c^2$ in the region below the convection zones in
    neighboring models as a function of envelope mass; both models
    have \teff=12,000~K, and have ML2/$\alpha$ values of 0.859 and
    0.885, respectively.  The dotted lines show the difference in the
    relevant quantities while the solid lines show an exponential fit
    to this difference. As the inset shows, these differences do not
    decay to zero with depth. }
\label{deltas}
\end{figure}

\section{Equilibrium Models}

The WD models used in these calculations were computed with the MESA
stellar evolution code
\citep{Paxton11,Paxton13,Paxton15,Paxton18,Paxton19}, and all the
models assumed a mass of $0.6\,M_\odot$ (the mean mass of WDs in our
sample is $0.62\,M_\odot$). Since the depth of the convection zone is
the key parameter in this study, we have used the mixing length
calibration of \citet{Tremblay15a} to set this
parameter. Specifically, we have used the value of
ML2/$\alpha_{\rm Schwa}$ from their Table~2 for the $\log\,g=8.0$
sequence interpolated to our \teff\ values of interest.

\begin{figure*}[t!]
  \centerline{
    \includegraphics[width=0.49\textwidth]{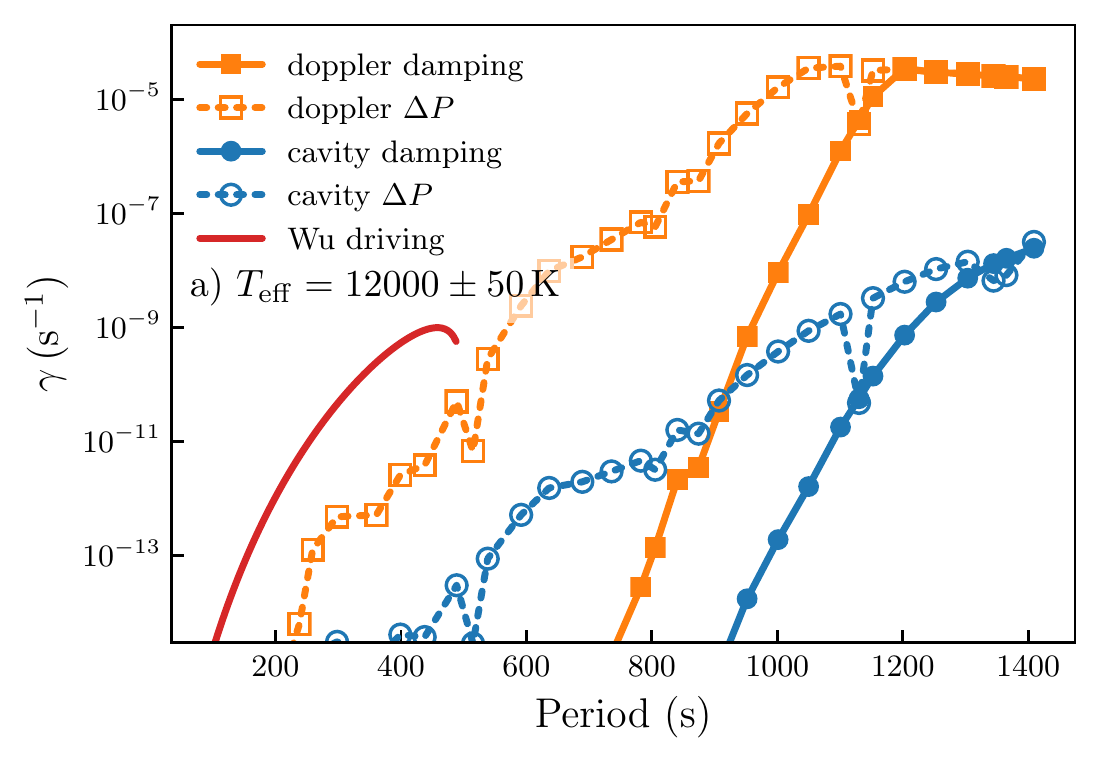}
\hfill
    \includegraphics[width=0.49\textwidth]{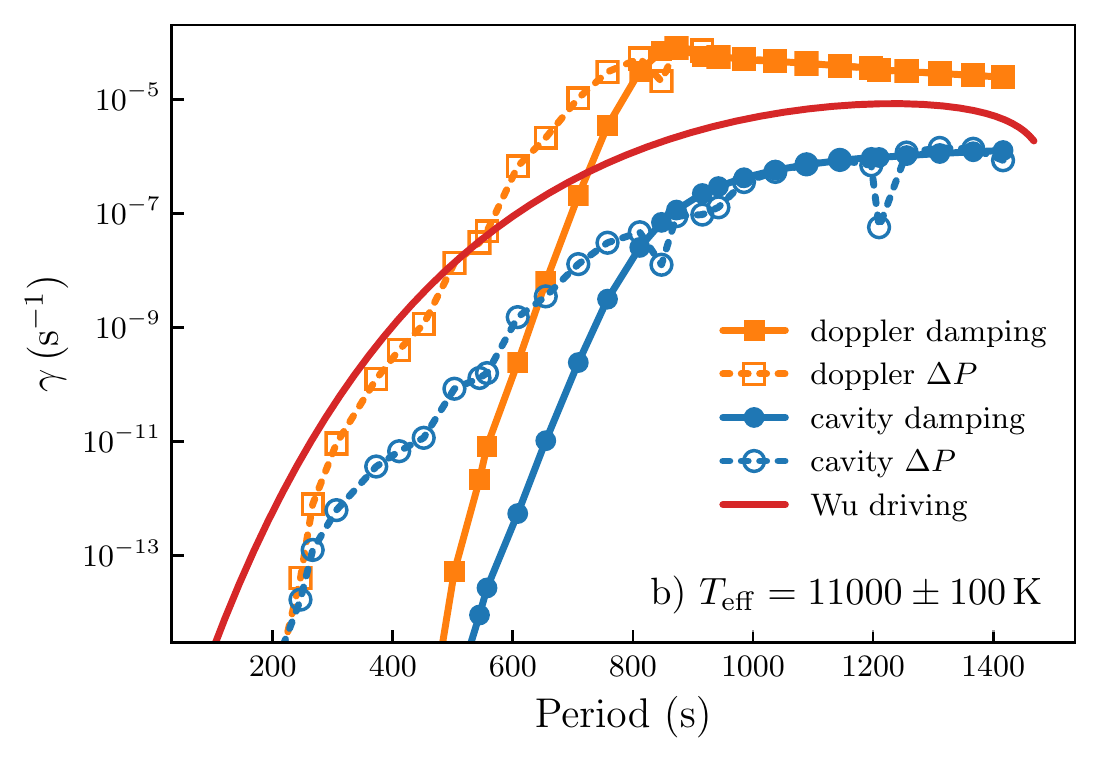}
  }
  \caption{A comparison of the finite-amplitude damping rates for
    $\ell=1$ modes due to a changing cavity size (blue curves) and
    those due to Doppler shifting of the reflected wave (orange
    curves); the curves with filled symbols employ a direct
    integration of the wavenumber (equation~\ref{dphi}) while those
    with open symbols use the period difference (equation~\ref{cav})
    with $\Delta P$ calculated as described in section~\ref{num}. We
    also show an estimate of the \emph{linear growth} rates of these
    modes (red curves) based on scaled results in \citet{Wu98}. Both
    sets of models have a mass of $0.6\,M_\odot$: those in a) have a
    \teff\ of 12,000~K with an assumed pulsation amplitude of 50~K,
    while those in b) have a \teff\ of 11,000~K with an assumed
    amplitude of $\pm$100~K.  }
  \label{damp}
\end{figure*}

To simulate the \teff\ changes due to finite amplitude pulsations, we
compute models with \teff\ values centered on the equilibrium state,
e.g., $\teff = 12,000 \pm 50\,$K. Since only the surface layers of the
model correspond to these perturbed \teff\ values, in a previous study
\citep{Montgomery15b} we spliced the outer portion of these models
onto the equilibrium model. This was justified by the fact that the
models rapidly converge with depth to nearly identical equilibrium
structures so that the splicing produces no visible numerical
artifacts. For the present study we adopt a different procedure. We
first compute the depth of the convection zone with the perturbed
\teff, say $\teff = 12,050\,$K. We then find the value of ML2/$\alpha$
that reproduces this depth for the equilibrium \teff\ of 12,000~K. We
now identify this model as the ``12,050~K'' model since it has a
convection zone depth that is identical to that of a ``true'' 12,050~K
model. This approach has the advantage that no splicing of models is
required.

\section{Numerical Issues}
\label{num}

For long-period modes whose outer turning points are very near the
base of the convection zone, all of our approaches yield very similar
results. That is, computing \dPhi\ from either the difference in
oscillation periods of two neighboring models (equation~\ref{cav}) or
using equation~\ref{dphi} to integrate the radial wavenumber in two
models yields the same result. For instance, this can be seen in
Figure~\ref{damp}b, in the agreement between the two sets of solid and
dotted curves for periods greater than $\sim 1000$s.

For shorter-period modes that have an outer turning point farther
beneath the convection zone, \dPhi\ depends on how quickly the
differences in model quantities decay with depth. Denoting the
differences between the two models (at constant mass coordinate) by
$\delta$, $\delta N^2/N^2$, $\delta c^2/c^2$, and
$\delta \omega_c^2/\omega_c^2$ initially decay exponentially with
depth. However, instead of decaying to zero, they asymptote to a level
of $\sim 10^{-6}$ (see inset in Figure~\ref{deltas}). Since these
models are designed to represent the same star at different times, the
internal structure of the models (which should not be strongly
affected by the changing surface convection zone) should be nearly
identical. We force this to be true in our analysis by fitting the
model differences with an exponential solution that decays to zero
with increasing depth. We then add these differences back to the
unperturbed model to create the new perturbed model, and these are the
models used in integrating \dPhi\ in equation~\ref{dphi}.  This leads
to a net decrease in the calculated damping rates of short-period
modes, which were dominated by small, unphysical model differences in
deeper layers.


In an asymptotic treatment, modes with outer turning points
sufficiently far beneath the perturbed region will be completely
unaffected by changes in the convection zone, and so will have zero
calculated damping. However, in a full calculation the modes are
global, and while they are evanescent in the region beneath the
convection zone, they do still sample it. For long-period modes that
propagate near the base of the convection zone, the period difference
$\Delta P$ between models is dominated by the difference in convection
zone depths, so we calculate $\Delta P$ as a direct difference of the
pulsation periods of corresponding modes in the two models
\citep{Montgomery15b}. For short-period modes, such period differences
are dominated by numerical artifacts from the deeper layers, so we
instead calculate the period difference via a variational approach
(see equation~14.19 of \citealt{Unno89} and equation~5.80 of
\citealt{JCD14}), i.e.,
\begin{equation}
  \Delta P = -\frac{\pi}{\omega^3} \, \frac{\int dM_r \, \delta N^2 \,
    \xi_r^2}{\int dM_r
    \left[\,\xi_r^2 + \ell (\ell+1) \,\xi_h^2 \, \right]},
\end{equation}
where $\xi_r$ and $\xi_h$ are the radial and horizontal displacements
of the spatial eigenfunction, and $\delta N^2$ is the difference in
the \bvfreq\ between the two models. We then use equations~\ref{cav},
\ref{domdop2}, \ref{dop2}, and \ref{gamma} to calculate the phase
shifts and damping associated with the period change of this mode.

\section{Results for $\ell=1$}
\label{res}

\begin{figure*}[t!]
  \centerline{
    \includegraphics[width=0.49\textwidth]{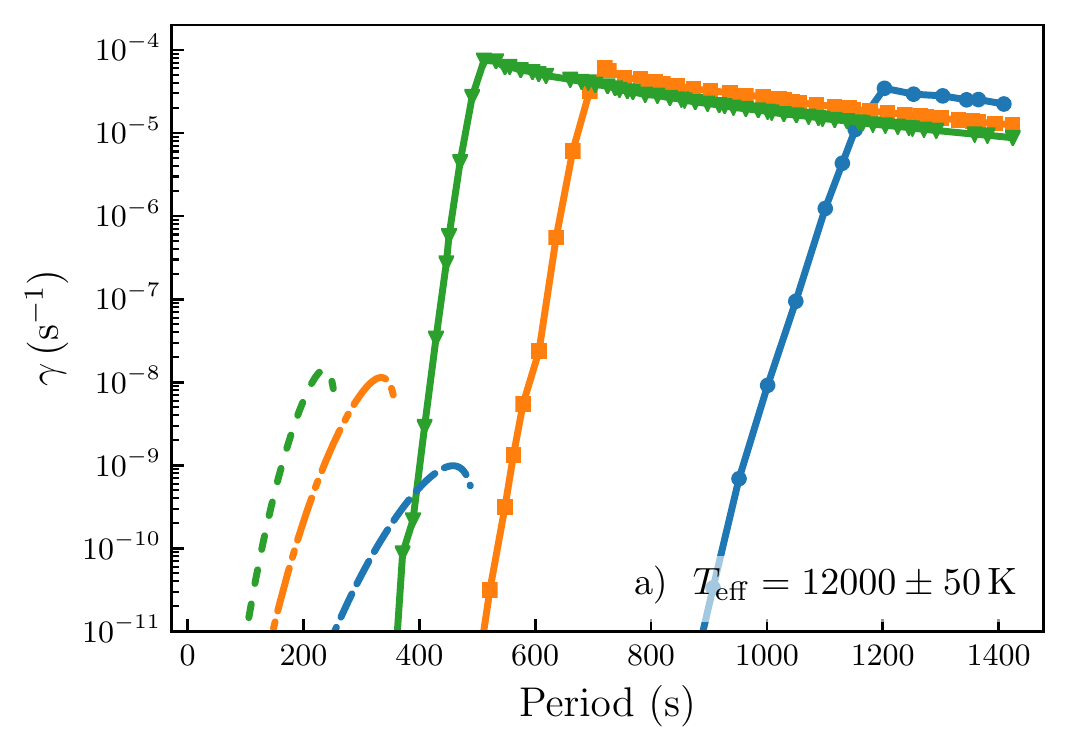}
\hfill
    \includegraphics[width=0.49\textwidth]{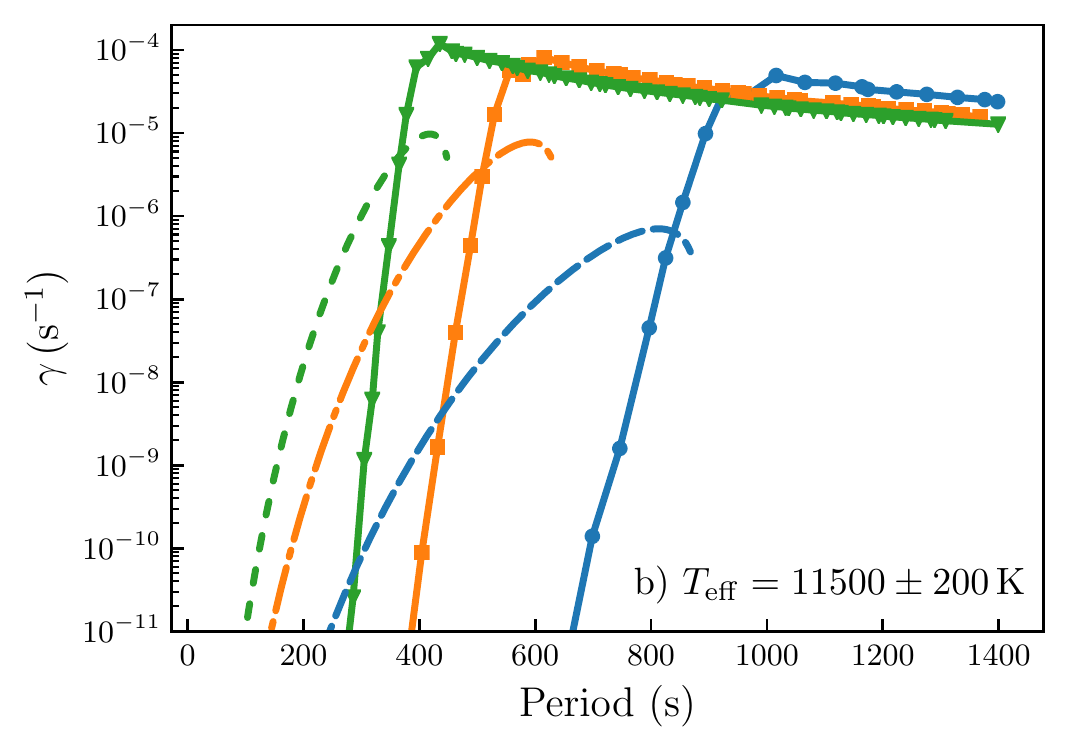}
}
  \centerline{
    \includegraphics[width=0.49\textwidth]{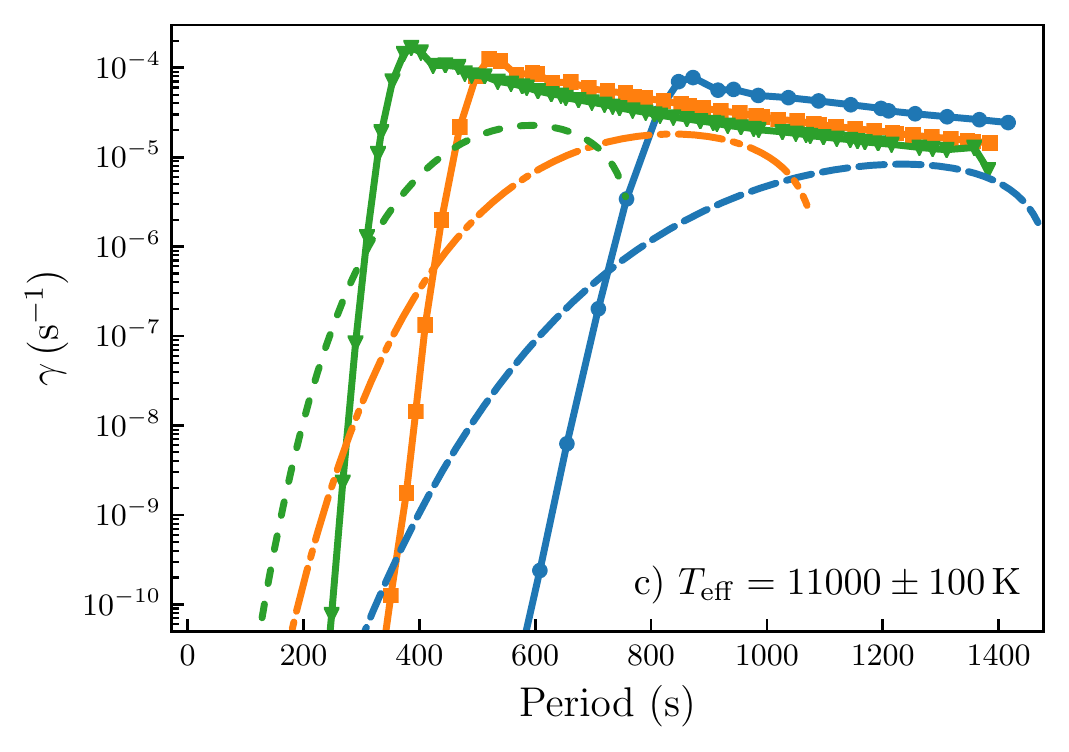}
\hfill
    \includegraphics[width=0.49\textwidth]{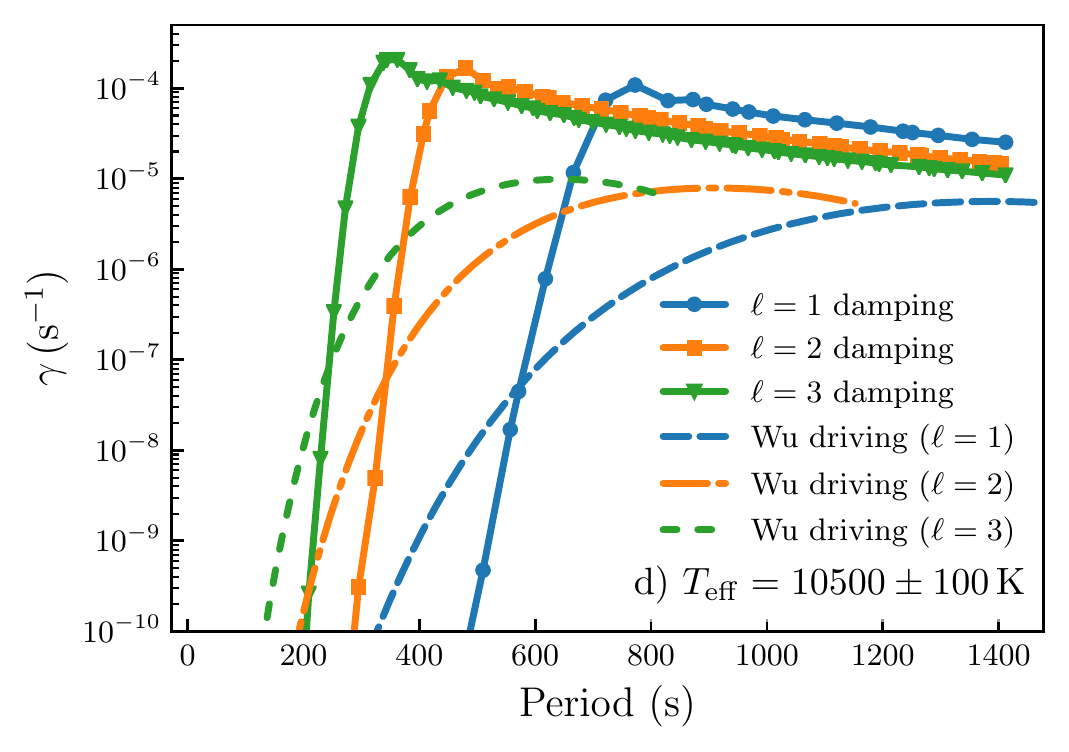}
  }
  \caption{The same as Figure~\ref{damp} showing the total damping
    results and estimated driving for a range of $\ell$ and \teff\
    values. Each plot contains results for $\ell=1$, 2, and 3 modes;
    panels (a)--(d) show results for \teff\ values of 12,000~K,
    11,500~K, 11,000~K, and 10,500~K, respectively. The $\pm$ values
    shown on each plot give the amplitude of the temperature
    perturbations assumed in calculating the nonlinear damping.  }
  \label{damp2}
\end{figure*}

Combining equations~\ref{cav}, \ref{dphi}, \ref{domdop2}, and
\ref{dop2} with equation~\ref{gamma}, we compute finite-amplitude
damping rates for $\ell=1$ pulsation modes. In Figures~\ref{damp}a,b
we show the damping rates using $0.6\,M_\odot$ WD models at two
different temperatures: $\teff = 12$,000~K for Figure~\ref{damp}a,
with assumed temperature excursions of $\pm 50$~K; and
$\teff = 11$,000~K for Figure~\ref{damp}b with assumed temperature
excursions of $\pm 100$~K.  After geometric cancellation, these cases
correspond to observed luminosity perturbations of $\sim$1\% and
$\sim$2\%, respectively. The blue curves with circular points show the
damping rate due to the changing size of the g-mode cavity, while the
orange curves with square points give the damping rate due to the
Doppler shifting of the reflected wave's frequency. In addition,
curves with filled symbols employ a direct integration of the
wavenumber (equation~\ref{dphi}) while those with open symbols use the
period difference (equation~\ref{cav}) with $\Delta P$ calculated as
described in section~\ref{num}. As can be seen, the two methods agree
well for long-period modes, e.g., the blue filled and unfilled symbols
in Figure~\ref{damp}b having $P > 1000$s.  The red curve in both plots
is an estimate of the \emph{linear growth} rates of these modes.
These estimates are based on calculations by \citet{Wu98}, as shown in
Figures~5.5--5.8 of her thesis (see also Figure~6 of \citealt{Wu99}
and Figure~5 of \citealt{Wu01a}).

The linear growth rates, by definition, do not depend on the amplitude
of the pulsations, while the nonlinear damping mechanisms presented
here do.\footnote{We note that a doubling of the \teff\ amplitude will
  approximately double the phase shifts, leading to a factor of four
  increase in the damping rates shown in Figures~\ref{damp} and
  \ref{damp2}.} As expected, we see that the damping due to the
Doppler shift of the mode's frequency is much larger than that due to
the variation in the size of its g-mode cavity, for both models and
sets of modes.  In Figure~\ref{damp}a, this driving is larger
  than the assumed total damping for all periods that are driven,
  using either method of calculating the damping, while in
  Figure~\ref{damp}b, the driving exceeds the damping for periods less
  than $\sim\,$650s when the damping is calculated by direct
  integration of the radial wavenumber (filled points), with the
  driving exceeding the damping for periods less than $\sim\,$500s if
  the period difference method is used (unfilled symbols).

\section{Damping as a Function of $\ell$ and \teff}

We next examine these results both as a function of $\ell$ and \teff.
In order to be conservative, in the remainder of this paper we use the
damping estimates computed by direct integration of the radial
wavenumber (e.g., filled symbols in Figures~\ref{damp}a,b), since
these appear to provide a lower limit for these rates. In
Figures~\ref{damp2}a--d, we show the finite amplitude total damping
rates as computed in the previous section for $\ell=1$, 2, and 3
modes. The different panels give the results for different values of
\teff; for context, we also plot estimates of the driving rates based
on Figures~5.5--5.8 from the thesis of \citet{Wu98}. In
Figure~\ref{damp2}a, the range of $\ell=1$ modes assumed to be driven
by convective driving is $P \sim $100--500~s (blue dashed curve),
which is plausible for a $\teff = 12000$~K model. For this case, we
see that the damping is essentially insignificant for the driven
modes. The same is true for $\ell=2$ and 3 modes.

Figure~\ref{damp2}b is a slightly cooler case ($\teff \sim 11$,500~K),
and the range of linearly driven $\ell=1$ modes is assumed to be
$P \sim $100--900~s, with ranges of 100--600~s and 100--450~s for
$\ell=2$ and 3, respectively. For the assumed 200~K amplitude, the
damping mechanism could now potentially affect the amplitudes of the
longest periods that are driven: 800~s for $\ell=1$, and 500~s and
350~s for $\ell=2$ and 3.

Finally, Figures~\ref{damp2}c,d show the trends at cooler
temperatures. As \teff\ decreases, the driving decreases relative to
the damping, implying smaller overall amplitudes. This has partially
been accounted for by a decrease in the assumed amplitude of the
pulsations, from $\pm 200$~K in Figure~\ref{damp2}b to $\pm 100$~K in
Figures~\ref{damp2}c,d.  For $\ell=1$, the period at which the driving
equals the damping decreases from 800~s at $\teff=$11,500~K to 650~s
and 550~s at $\teff=$11,000~K and 10,500~K, respectively. A similar
trend can be seen for $\ell=2$ and 3 modes. Overall, these results
imply that both the amplitudes seen in cooler models and their maximum
periods should decrease.  While the decrease in overall amplitude near
the observed red edge has been previously noted \citep{Mukadam06b},
any observational decrease in maximum period with \teff\ is less
dramatic and has not been conclusively identified.

\begin{figure}[t!]
  \centerline{\includegraphics[width=1.0\columnwidth]{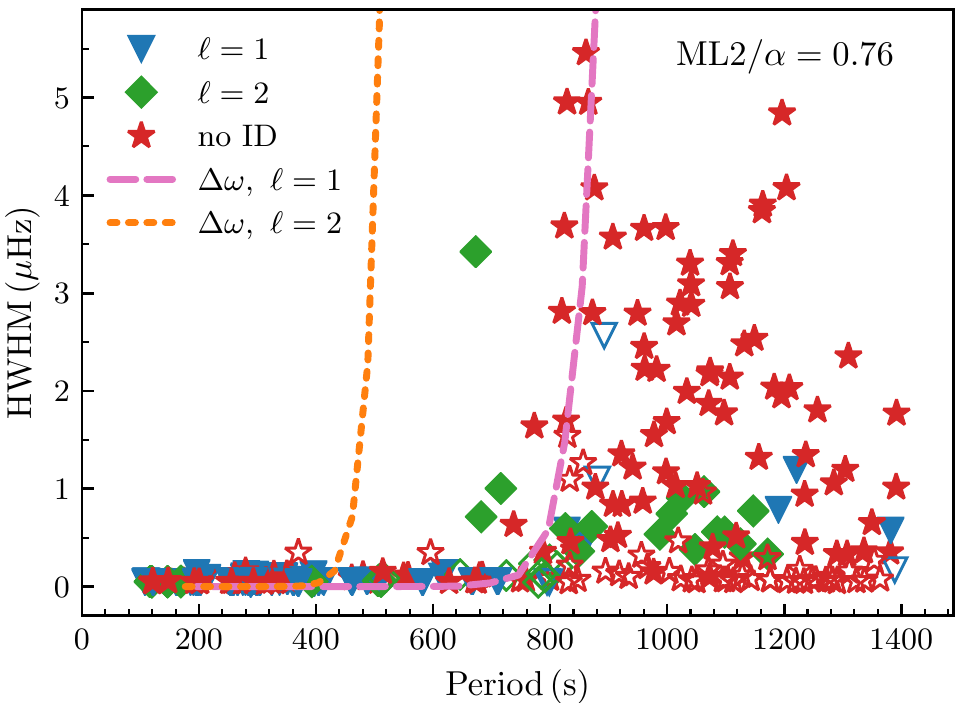}}
  \caption{The same as Figure~\ref{predict} but with a calculation of
    the expected mode widths for $\ell=1$ (dashed magenta curve) and
    $\ell=2$ (dotted orange curve) modes superimposed on the
    data. This calculation assumes a model with $\teff = 11,500\,$K,
    $\logg = 8.0$, $\alpha = 0.76$, with the \teff\ variations due to
    pulsation taken to be $\pm 200\,$K.  }
  \label{m_width}
\end{figure}

\section{Mode Widths}

Due to the nature of the stochastic excitation mechanism, the
theoretical damping rates of solar-like pulsators provide a prediction
for the observed Fourier widths of the modes. While this connection is
less clear for pulsators with modes that are linearly unstable (such
as the DAVs), we will attempt to calculate mode widths in the context
of the damping mechanism presented above.\footnote{A mode that is
  linearly driven is usually assumed to have grown in amplitude to the
  point that further growth is limited by some nonlinear process,
  leading to a stable limit cycle. At this point its amplitude and
  phase are nearly constant in time, resulting in mode widths that are
  \emph{much} smaller than those given by the calculated driving and
  damping rates.}

Equation~\ref{domdop} provides an estimate of the frequency change
produced by a single interaction of a wave with the convection
zone. If we interpret this frequency shift as an estimate of the width
of the Fourier peak of the mode, we can calculate the expected widths
for the pulsation modes in a given stellar model.

We show the results of such a calculation in Figure~\ref{m_width}.
The dashed magenta curve shows the frequency width for $\ell=1$ modes
calculated in this way, while the dotted orange curve shows the width
for $\ell=2$ modes. These calculations are based on the equilibrium
model of Figure~\ref{damp2}b ($\teff = 11,500\,$K, $\logg = 8.0$,
$\rm ML2/\alpha = 0.76$, with \teff\ variations of $\pm 200\,$K). We
see that there is a steep rise in the $\ell=1$ predicted widths at
$\sim 800$~s, nominally coinciding with the observed rise seen in the
data. We also note that the $\ell=2$ modes with $P \sim 700\,$s would
not be expected to have large widths if they were $\ell=1$ modes, but
are easily accounted for as $\ell=2$ modes. While certainly not the
final word, this calculation provides a partial explanation for the
observed rise in mode widths with increasing period.

\section{Discussion}

In the preceding sections we have shown that the modes with observed
Fourier widths greater than 0.3~$\mu$Hz have longer periods
($P \gtrsim 800$~s), and that in most cases these modes are calculated
to propagate all the way to the base of the surface convection zone in
pulsating DA WDs. We have proposed mechanisms through which the
convection zone can cause a lack of coherence in these modes.  We find
the dominant mechanism to be the Doppler shift of the mode frequency
as it reflects off the time-dependent outer turning point.
Preliminary results are that this mechanism is stronger near the
observed red edge of the instability strip and may be important for
the observed reduction in mode amplitudes there.



The phase shifts are a finite amplitude effect that naturally lead to
damping. For hotter models ($\teff \ga 12,000$~K), this damping
  is quite small and likely unimportant.  On the other hand, for
  cooler models ($\teff < 12,000$~K), this damping can be significant
  for longer period modes. The DAVs known to have stable pulsations
coherent over a time period of years have short period pulsations
  ($P \la 400\,$s) and are near the blue edge of the instability
strip. From Figure~\ref{damp}a, we see that the proposed damping
  mechanism is quite small for these modes, which could be part of the
  reason these modes exhibit such extreme coherence. For cooler
  models, the damping is still small for the short-period modes, so
  the possibility remains that cooler pulsators could also harbor
very stable pulsations in the form of low-period modes ($P <
300$~s). These modes could have propagation regions sufficiently
distant from the convection zone that they would still have extreme
stability. Such modes, if they exist, could be used to
asteroseismically trace secular evolution of cooler DAVs or expand the
search for planets around white dwarfs with pulsation timing
variations. As far as we are aware, no systematic search for such
stable modes in cooler DAVs has been made, though some short-period
modes do appear stable in cool DAVs over the span of months in
\emph{Kepler/K2} observations.

We note that the calculations presented here could be improved in two
respects.  First, the convective driving rates based on the formalism
of either \citet{Wu99} or \citet{Dupret04} and \citet{VanGrootel12}
could be calculated for the stellar models employed here, leading to a
consistent set of driving and nonlinear damping rates. In addition,
the calculation of the difference in structure of neighboring models
could be improved. Although the differences in their outer layers are
dominated by their different convection zone depths, which our current
calculations take into account, the differences deeper in the models
will be directly due to the pulsation modes themselves. Thus,
realistic eigenfunctions should be used to calculate the instantaneous
structure of these models, which are then used in the subsequent
analysis.

Finally, many DAVs observed with the \emph{Kepler} and \emph{K2}
missions undergo outbursts, increases in average brightness of
10--40\% that can typically last from 5 to 15 hours
\citep{Bell15a,Hermes15b,Bell17a}. While the best-known theory for
this process involves a resonant transfer of energy from a driven
parent mode to damped daughter modes as an amplitude threshold is
reached \citep{Luan18,Wu01a}, we speculate that the mechanism we have
proposed involving phase shifts of reflected modes could be relevant
as well. It is possible that some pulsators reach an amplitude
threshold in which there is a slight increase in the damping, and this
increased damping leads to a slight heating of the surface
layers. This in turn causes a net thinning of the convection zone,
leading to larger phase shift mis-matches, which leads to further
damping, and the cycle reinforces itself. In addition, as the surface
convection zone becomes thinner, radiative damping may play an
important role in removing energy from the high overtone modes
\citep{Luan18}.

Nonlinear resonant energy transfer also appears to modify the
frequency and amplitudes of modes on timescales shorter than expected
from secular evolution, and this has been observed in a number of
pulsating white dwarfs with long-baseline observations
\citep[e.g.,][]{Dalessio13,Zong16}. However, the frequency changes of
these effects are considerably smaller than those observed in the
broadened mode line widths discussed here. Additionally, many of the
DAVs with frequency changes incompatible with secular evolution occur
in hotter stars with short-period ($< 300\,$s) pulsations that should
be relatively unaffected by the convection zone (e.g.,
\citealt{Hermes13b}).

\section{Conclusions}

We present a mechanism that may be relevant for the limitation of
pulsation amplitudes in white dwarf stars for modes above a threshold
period. As the convection zone changes depth during the pulsation
cycle, the condition for coherent reflection of the outgoing traveling
wave is slightly violated. In effect, this causes the amplitude of the
mode (viewed as the superposition of inward- and outward-propagating
components) to decrease, leading to damping. This mechanism should be
present at some level in all pulsating WDs, and should be larger near
the red edge of the DAV instability strip.  In addition, this
mechanism could possibly be relevant for other g-mode pulsators with
surface convection zones (e.g., Gamma Doradus stars), or even
large-amplitude pulsators such as high-amplitude delta Scuti stars
(HADs) or RR Lyrae stars.



\section*{Acknowledgements}

MHM and DEW acknowledge support from the United States Department of
Energy under grant DE-SC0010623 and the NSF grant AST 1707419.  MHM,
DEW, and BHD acknowledge support from the Wootton Center for
Astrophysical Plasma Properties under the United States Department of
Energy collaborative agreement DE-NA0003843. JJH acknowledges support
from NASA K2 Cycle 5 grant 80NSSC18K0387 and K2 Cycle 6 grant
80NSSC19K0162. KJB is supported by an NSF Astronomy and Astrophysics
Postdoctoral Fellowship under award AST-1903828.

\software{MESA \citep{Paxton11,Paxton13,Paxton15,Paxton18,Paxton19},
  Scipy \citep{Jones01}, Numpy \citep{Oliphant06}, Matplotlib
  \citep{Hunter07}}




\bibliography{index_f.bib}
\bibliographystyle{aasjournal}


\end{document}